# Development and Evaluation Study of Intelligent Cockpit in the Age of Large Models

Jun Ma[1,2,*], Meng Wang[1], Jinhui Pang[3], Haofen Wang[1], Xuejing Feng[1], Zhipeng Hu[1], Zhenyu Yang[1], Mingyang Guo[1], Zhenming Liu[1], Junwei Wang[1], Siyi Lu[1], Zhiming Gou[1]

1. XAI LAB, College of Design and Innovation, Tongji University, Shanghai 200092, China
2. School of Automotive Studies, Tongji University, Shanghai 201804, China
3. School of Computer Science & Technology, Beijing Institute of Technology, Beijing 100081, China

*Corresponding author. XAI LAB, College of Design and Innovation, Tongji University, Shanghai 200092, China. E-mail address: jun_ma@tongji.edu.cn.

**Abstract**: The development of Artificial Intelligence (AI) Large Models has a great impact on the application development of automotive Intelligent cockpit. The fusion development of Intelligent Cockpit and Large Models has become a new growth point of user experience in the industry, which also creates problems for related scholars, practitioners and users in terms of their understanding and evaluation of the user experience and the capability characteristics of the Intelligent Cockpit Large Models (ICLM). This paper aims to analyse the current situation of Intelligent cockpit, large model and AI Agent, to reveal the key of application research focuses on the integration of Intelligent Cockpit and Large Models, and to put forward a necessary limitation for the subsequent development of an evaluation system for the capability of automotive ICLM and user experience. The evaluation system, P-CAFE, proposed in this paper mainly proposes five dimensions of perception, cognition, action, feedback and evolution as the first-level indicators from the domains of cognitive architecture, user experience, and capability characteristics of large models, and many second-level indicators to satisfy the current status of the application and research focuses are selected. After expert evaluation, the weights of the indicators were determined, and the indicator system of P-CAFE was established. Finally, a complete evaluation method was constructed based on Fuzzy Hierarchical Analysis. It will lay a solid foundation for the application and evaluation of the automotive ICLM, and provide a reference for the development and improvement of the future ICLM.

# 1. Research and Application of Intelligent Cockpit and Large Models

## 1.1 Intelligent Cockpit: An Overview

An intelligent cockpit refers to a vehicle's driving space that integrates multiple smart technologies and human-machine interaction methods. The development of intelligent cockpits has gone through three significant phases:

**First Generation**: This phase primarily involved transitioning traditional mechanical equipment to digital devices, integrating various in-car systems such as music players and navigation, which significantly improved the driving and passenger experience.

**Second Generation**: By integrating in-car devices like refrigerators, televisions, and seats with massage and heating functions, the second generation focused on enhancing the cockpit's comfort and entertainment value. Basic voice control systems also became standard features during this period.

**Third Generation**: In recent years, rapid technological advancement and evolving consumer demands have spurred multiple innovations in the intelligent cockpit space. The adoption of centralized automotive architectures, IoT (Internet of Things), and Cloud Services has significantly driven the development and application of intelligent cockpits. The trend is shifting toward spatial computing and full-scene intelligence, turning the cockpit into more than just a space for driving and riding; it is becoming a "third space" for leisure, entertainment, and work. Intelligent cockpits seamlessly connect with smartphones, personal computers, and smart home devices, offering users convenience and smart experiences similar to being at home.

**Key Technologies of Intelligent Cockpits**

As a platform for integrating various emerging technologies in intelligent vehicles, the intelligent cockpit is characterized by its wide technical scope and strong scene expansion capabilities, involving numerous academic fields. Key technologies include, such as software, hardware technology and development technology in terms of overall technology framework (as shown in Figure 1) [1]:

**Environmental Sensing Technology**: This uses cameras, microphones, radar, and other sensors to monitor the vehicle's internal and external environment in real-time, providing data to support intelligent decision-making.

**Vehicle-to-Everything (V2X) Technology**: Through in-car communication modules, vehicles can exchange information with the internet, other vehicles, and infrastructure, improving traffic efficiency and safety.

**Artificial Intelligence Algorithms**: AI is applied in areas such as route planning, proactive recommendations, and emotion recognition, enhancing the intelligence level of the cockpit.

**Multimodal Human-Machine Interaction**: This includes interfaces like car-machine interaction, voice recognition, gesture control, and facial recognition, allowing for more natural and intuitive interactions.

**Current Market Status of Intelligent Cockpit Products**

Driven by cutting-edge technologies such as AI, machine learning, and V2X communication, the development and application of intelligent cockpits have made remarkable progress. AI and machine learning algorithms play a crucial role in enhancing cockpit functionality, significantly improving interactivity, adaptability, and the overall user experience. Advanced driver monitoring systems enable AI to analyze facial expressions, eye movement, and physiological signals to assess the driver's emotional and cognitive state, allowing the vehicle to issue timely warnings or make adjustments to improve safety and comfort [2]. AI also supports predictive analytics, allowing intelligent cockpits to anticipate user needs and preferences, such as adjusting climate controls based on historical data and real-time conditions or suggesting optimal routes [3]. Additionally, AI-driven voice assistants and gesture recognition systems further improve hands-free control, minimizing driver distractions and enhancing convenience [4]. Ergonomics and user experience remain at the core of cockpit development, ensuring that interfaces are user-friendly and that overall comfort is prioritized [4].

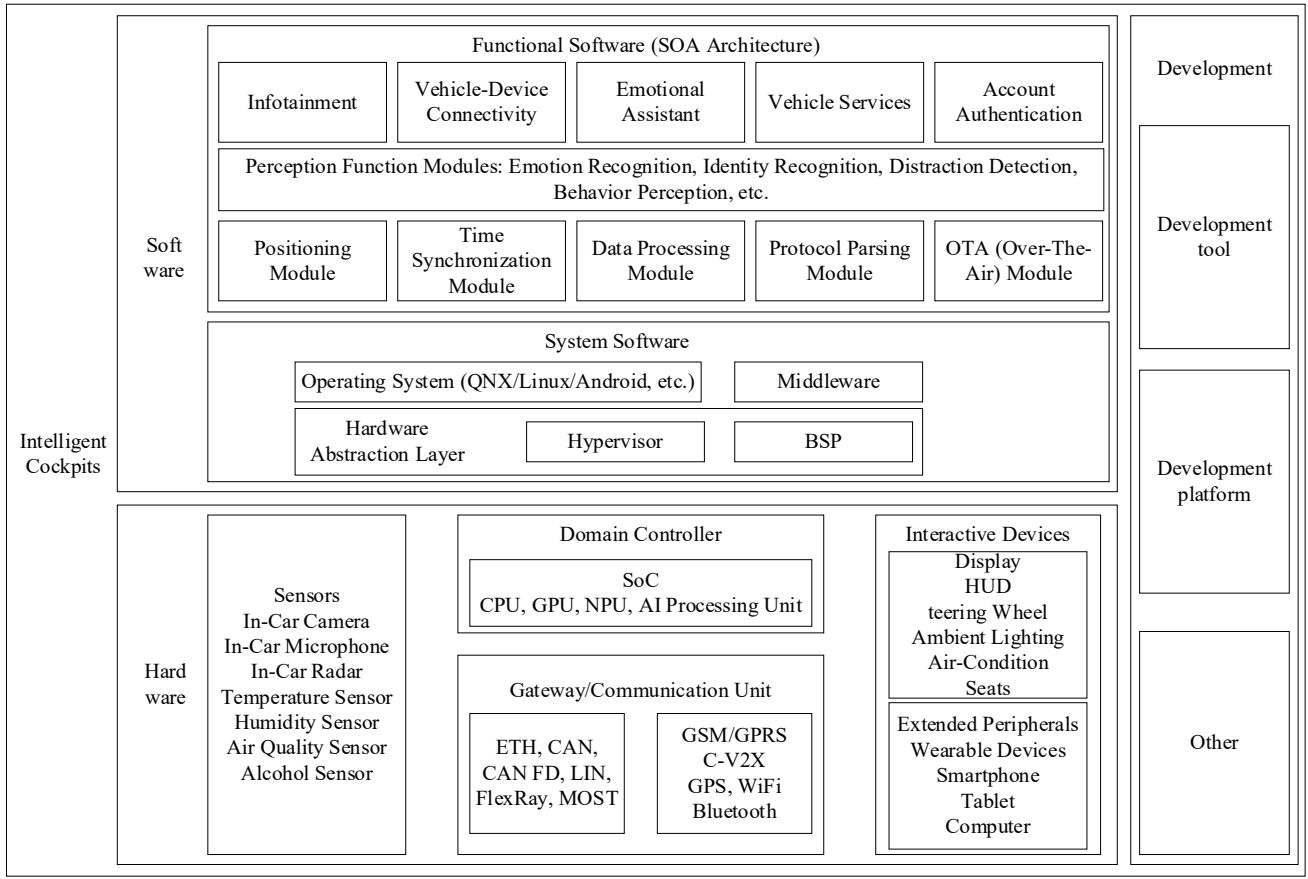

Figure 1 Technical Architecture of Intelligent Cockpits

By 2023, the penetration rate of intelligent cockpits had exceeded 80%. Major automakers like Tesla, Geely, and GAC have launched cockpit products equipped with advanced technology, such as high-definition central control displays and heads-up displays (HUD). These vehicles are also integrated with intelligent voice assistants. New energy vehicle manufacturers such as Li Auto, XPeng, and NIO place particular emphasis on the development of intelligent cockpits and user experience, promoting them as key selling points. Consumer research reports show that the importance of intelligent cockpits in car-buying decisions continues to grow.

As the "brain" of autonomous driving, the Intelligent Cockpit has traditionally been based on deep-learning AI models. The rapid advancement and implementation of large models are seen as the next generation of this "brain", as shown in Table 1. The trend for intelligent cockpits is to move toward greater intelligence and improved interactive experiences, which not only enhance driving comfort and convenience but also provide users with new lifestyle and workstyle options.

**1.2 Development and Application of Artificial Intelligence and Large Language Models**

Artificial Intelligence (AI) refers to technology that enables machines to replicate human cognitive functions such as recognition, analysis, and decision-making. The essentiality of AI is simulating human consciousness and thought processes. AI is a multidisciplinary field, combining computer science, physiology, philosophy, and more, to create systems that mimic human intelligence [5]. The application of AI in the automotive industry has moved from concept to reality, becoming one of the core drivers of innovation in the sector.

Over the past five years, in-vehicle AI technology has developed rapidly. In 2019, the commercialization of 5G made vehicle-to-everything (V2X) technology possible, significantly expanding the scope for AI applications in vehicles. By 2020, with the maturity of deep learning techniques, in-car AI began to enable more precise and intelligent control across perception, decision-making, and execution layers. In 2021, government policies and rising market demand led to a significant increase in the adoption of intelligent vehicles. By 2022, breakthroughs in AI chip technology provided more powerful computational capabilities for in-vehicle AI. By 2023, AI applications in cars spanned everything from advanced driver-assistance systems (ADAS) to fully autonomous driving.

Table 1 Current Status of Large Model Development in Intelligent Cockpits

| Company | Vehicle Model | Large Model Developed/Implemented |
| --- | --- | --- |
| XPeng Motors | X9 | AI Tianji / XGPT Lingxi |
| NIO | 2024 EC6 | NOMI GPT / End-Cloud Multimodal Model |
| Li Auto | MEGA | Mind GPT |
| Huawei | Huawei Qian Kun | Qianwu Engine Model |
| Jiyue Motors | Jiyue 01 | Baidu-Wenxin Yiyan |
| Seres | AITO M9 | Huawei-Pangu |
| Neta | Neta L | NETA GPT |
| Xiaomi | SU7 | MiLM-1.3B |
| GAC Aion | Hyper GT | GAC AI Model |
| IM Motors | IM LS6 | IM Generative Model |
| Geely Zeekr | Zeekr 007 | Kr GPT |
| Geely Wingle | Wingle L380 | SenseTime Ririxin Model |
| Geely | Geely Galaxy L6 (OTA) | Xingrui |
| Changan Deepal | Deepal S7i | Deepal GPT |
| Chery | Xingji ES | iFlytek-Spark Large Model LION AI |
| Changan | E07, Yida | Baidu-Wenxin Yiyan |
| BYD | N/A | Xuji Model |
| Great Wall | Wey Lanshan DHT-PHEV | Space GPT, Haomo DriveGPT |
| Volkswagen | ID.7, ID.4, ID.5, etc. | ChatGPT |
| Cadillac (SAIC GM) | IQ Aoge / Rui Ge | Baidu |
| Dongfeng Lantu | - | VCOS GPT Model |
| GAC | - | GAC AI |
| FAW Hongqi | - | Baidu-Wenxin Yiyan |
| BMW | - | Amazon Alexa Large Language Model |
| Mercedes-Benz | - | ChatGPT |

The application of AI in the field of intelligent driving is mainly reflected in ADAS and autonomous driving technology [6]. Another key application area is the intelligent cockpit. By integrating speech recognition and natural language processing (NLP) technologies, in-car systems can understand and respond to driver's commands and questions, enabling more natural and efficient interactions. AI is driving significant changes in the interaction ecosystem of intelligent cockpits, moving towards more natural, seamless user experiences. From smart driving to Intelligent cockpits, AI is continuously driving innovation in the automotive industry. With ongoing advancements, AI's role in the automotive sector will become even more widespread and profound.

**Development and Current Status of Large (Language) Models**

Large models refer to deep neural networks with billions of parameters, demonstrating intelligence levels comparable to humans in areas such as knowledge comprehension, reasoning, and prediction. These models can be classified into multimodal models, which process text, speech, images, and video data, and large language models (LLMs), trained on vast amounts of text data [7]. The emergence of LLMs, such as GPT-4, is regarded as a breakthrough in AI development.

Modern LLMs have also demonstrated impressive multimodal capabilities, handling tasks such as natural language understanding and generation. For example, Transformer-based models are capable of implementing the tasks for generating text, translating, and abstracting; Multimodal content understanding and generation, e.g., based on the Diffusion model, can generate images, audio and video based on textual descriptions. In practice, there is little distinction between large models and large language models, as both now refer to systems capable of understanding, generating, and interacting with multimodal data based on human knowledge.

Applications such as ChatGPT and Stable Diffusion, built on large models, have attracted hundreds of millions of users within a year. This highlights the significant potential for these models to transform the intelligent cockpit in vehicles. Overall, the integration of large models into vehicles is becoming an inevitable trend, accelerating the development of automotive intelligence and enhancing human-vehicle interactions for an improved driving experience.

**Impact of Large Models on Intelligent Cockpits**

LLM's capabilities in NLP have revolutionised in-vehicle systems, particularly in areas such as intelligent voice assistants, natural language understanding, driver behaviour analysis, and personalized services.

LLMs have significantly enhanced the natural interaction capabilities of voice assistants. By leveraging LLMs, these systems can improve the accuracy and fluidity of speech recognition and natural language processing, enabling more natural conversations with drivers. This allows for voice control of various vehicle functions and provides a smoother, more personalized interaction. For instance, based on historical user data and real-time conditions, the system can offer tailored recommendations, such as suggesting routes, music, or restaurants based on users' daily habits and preferences.

In terms of natural language understanding, LLMs enable in-car systems to comprehend complex commands and queries, supporting multi-turn conversations and contextual understanding. This has enhanced the interactivity and functionality of operating systems, allowing users to perform more complex tasks through voice commands. Advanced features like logical reasoning, strategy planning, and knowledge-based Q&A have further upgraded vehicle operating systems, making voice interactions one of the primary modes of operation and prompting adjustments in system architecture.

Moreover, LLMs can analyse driver behaviour to deliver personalized services. By evaluating a driver's language and behavioural patterns, the AI system can offer tailored driving experiences, such as adjusting seat settings, music playlists, and navigation preferences based on user habits [8].

On-device deployment of large models helps protect data privacy. Currently, only a few manufacturers, such as NIO and Xiaomi, are planning on-device deployments for intelligent cockpits, while most still rely on cloud-based systems. However, according to the application trend in the mobile phone field, the End- Cloud combined model deployment scheme is the trend, a hybrid model of edge and cloud deployment is expected to become mainstream. Companies like Xiaomi (1.3B model), Google (7B model), and Apple (3B model) are working on compressing model parameters to enable on-device deployment while maintaining the model's capabilities.

**1.3 Research Status of LLM-based Agent**

"Agent AI" refers to interactive systems capable of processing visual stimuli, language inputs, and other environment-based data to generate meaningful actions [9]. LLM-based agents (hereafter referred to as LLM-Agent) use large language models to facilitate intelligent interaction and have shown significant potential in fields such as natural language processing, healthcare, automotive, finance, and education.

In the automotive sector, LLM-Agents are mainly applied to environmental perception, decision planning, and human-vehicle interaction. Research focuses on improving perception capabilities in various conditions through multi-sensor fusion, optimizing decision algorithms for quick and accurate responses in complex traffic scenarios, and developing more natural and intelligent human-vehicle interaction systems, including voice assistants and gesture recognition. In NLP, LLM-Agent has made notable advancements in machine translation, text generation, sentiment analysis, and dialogue systems. Research emphasizes enhancing models' understanding of complex semantics and context, improving multilingual capabilities, increasing the accuracy of sentiment and intent recognition, and advancing context processing in dialogue systems.

Although LLM-Agent demonstrates broad application potential, there are still many challenges to overcome. Data bias and fairness are significant issues in NLP, where biased training data can lead to prejudiced outputs. In fields such as automotive and healthcare, model transparency and explainability remain key concerns, as the "black box" nature of current models undermines trust in decision-making processes. Data privacy and security are critical challenges across all fields, ensuring the protection of user data during use and storage. Real-time processing and the computational demands of LLM-Agent are especially crucial in automotive and fintech sectors, where large volumes of data need to be processed quickly to make time-sensitive decisions. The reliability and safety of these models in extreme weather conditions or complex traffic scenarios also require further validation. Human-computer interaction and education sectors face personalization and adaptability challenges, where research is needed to tailor large models to meet individual user needs effectively. In addition, enhancing the naturalness and user experience of the interaction system, as well as realising multimodal interaction in the in-car environment, are also research focuses. Lastly, ethical and legal issues are essential across all sectors, particularly in areas like liability in autonomous vehicles and data privacy in education, which require robust societal and legal frameworks to resolve.

### 1.4 Current Research on LLM-Agent in Automotive Intelligent Systems

In the study of LLM-Agent within automotive intelligent systems, the LLM-Agent often serves as an intermediary bridge, connecting the driver and the vehicle to form a three-way interaction. This interaction encompasses three main dimensions: information exchange, decision support, and safety monitoring. In terms of information exchange, the LLM-Agent facilitates communication with the driver through natural language and converts the driver's commands into vehicle operation instructions. For decision support, it analyzes real-time information provided by the driver and the vehicle's status to offer navigation guidance, route selection, and driving behavior adjustments. In safety monitoring, the LLM-Agent can track the driver's state, such as fatigue or distractions, issuing timely warnings or suggesting rest breaks, thereby enhancing driving safety.

Research in this field primarily focuses on five areas: natural language understanding and generation, multimodal interaction, emotion and intent recognition, decision support, and safety monitoring and driving assistance. Natural language understanding and generation involve the LLM-Agent efficiently comprehending the driver's commands and responding contextually. In terms of multimodal interaction, the LLM-Agent integrates multiple modes of input, such as voice and visual information, to provide a more comprehensive interactive experience in complex driving scenarios. For example, it can respond to gestures or facial expressions captured by cameras while interpreting voice commands. Emotion and intent recognition involve analyzing the driver's speech and text to identify their emotional state and intent, providing personalized and humanized services. In decision support, the LLM-Agent processes and analyzes large amounts of data in real time, such as traffic conditions, weather data, and vehicle status, offering suggestions for navigation, route planning, and driving behavior adjustments. In complex driving environments, the Agent can recommend optimal routes based on traffic conditions and the driver's current state or alert them to potential hazards. Furthermore, the Agent can learn and adapt to different user preferences, offering personalized services by adjusting music, temperature, and seat settings based on the driver's history and current mood, enhancing the overall user experience. In safety monitoring and driving assistance, the LLM-Agent continuously monitors the driver's condition, detecting signs of fatigue or distraction, and issues warnings or suggests rest breaks to reduce the risk of accidents.

Through its advanced capabilities in natural language understanding, multimodal interaction, emotion and intent recognition, decision support, and safety monitoring, the LLM-Agent has shown great potential and broad application prospects in automotive intelligent systems. These technological advancements not only improve driving safety and user experience but also provide new ideas and directions for the development of intelligent cockpits.

## 2. Development of Intelligent Cockpits Integrated with Large Model

At present, many users criticize the human-machine interaction capabilities in cars. However, large language models and other generative models applied in intelligent cockpit and vehicle networking scenarios have demonstrated great potential, leading many automakers to consider them a strategic reserve technology. Large language models offer Original Equipment Manufacturers (OEM) new approaches by leveraging deep learning and language generation to achieve more natural and open-ended human-machine interactions. This represents a significant improvement in the personalization, emotional engagement, and freedom of interaction in intelligent cockpits [10].

### 2.1 Current Research on the Integration of Large Model and Agents in Intelligent Cockpits

The intelligent cockpit is a key component of modern vehicle human-machine interaction, aiming to provide drivers and passengers with a more convenient, safe, and comfortable driving experience. In recent years, with the advancement of LLM-Agent technology, the interactive capabilities of intelligent cockpits have significantly improved. This section provides an overview of current research on the application of LLM-Agent in intelligent cockpits, exploring progress in natural language interaction, emotion and intent recognition, multimodal interaction, and intelligent decision-making.

In natural language interaction, large language models enable intelligent cockpits to deliver more natural and fluid voice interactions. LLM-driven voice assistants can understand complex natural language commands and provide services such as navigation, entertainment control, and information queries. For example, drivers can use voice commands to adjust the air conditioning, play music, or check the weather. LLM-based dialogue systems support multi-turn conversations, maintaining contextual consistency for a more intelligent interactive experience. For instance, drivers can engage in continuous dialogue with the system to modify travel plans or receive real-time traffic updates.

In decision support, the intelligent cockpit's ability to provide decision-making support based on multimodal data is a critical function, with the LLM-Agent playing a key role. By analyzing the driving environment and driver status in real-time, the cockpit can offer route planning, danger warnings, and driving advice. For example, in complex road conditions, the system may recommend the best route. It can also make personalized suggestions based on the driver's habits and preferences, such as adjusting seat positions, air conditioning temperatures, or music playlists.

In emotion and intent recognition, the intelligent cockpit processes multimodal information to achieve emotion and intent recognition [11]. Using emotion recognition technology, the LLM-Agent can analyze the driver's speech and facial expressions to identify emotional states and respond accordingly. By analyzing the driver's voice content and tone, it can detect emotions such as anger, fatigue, or anxiety. Based on this analysis, the system can offer personalized feedback, such as recommending a rest break or playing calming music when signs of fatigue are detected.

In multimodal integration, the intelligent cockpit combines data from various sensors, including voice, video, and vehicle sensors, to achieve comprehensive environmental awareness and interaction capabilities. For example, by capturing the driver's facial expressions with cameras and combining them with voice commands, the system can better understand the driver's intent. Integrating onboard sensor data, such as speed, acceleration, and direction, the LLM-Agent can monitor changes in the vehicle and surroundings in real-time, providing more accurate driving assistance.

The integration of LLM-Agent technology into intelligent cockpits has greatly enhanced the naturalness and intelligence of human-machine interactions. Although challenges remain, this integration shows immense potential for improving both driving experience and safety, warranting further research and exploration. However, research on user feedback regarding the capabilities and experiences of large models in intelligent cockpits is still lacking. Therefore, a comprehensive and objective evaluation system is needed to help stakeholders (such as users and automakers) better understand and assess the overall and detailed capabilities of large models in intelligent cockpits, serving as a foundation for future improvements.

## 2.2 Limitations in Establishing an Evaluation System for ICLM

The cockpit is a highly important and complex human-machine interaction system, responsible for enabling all interaction functions, controlling the entire vehicle, and ensuring the driver's physiological and psychological safety within a limited space. The design quality of the cockpit directly impacts the driver's experience, driving efficiency, and safety. Therefore, each type of cockpit requires a thorough evaluation after its design is completed to verify whether it meets the required standards.

Traditionally, evaluation methods involve expert panels composed of industry professionals and drivers. However, these methods tend to rely on subjective judgments and often lack objective data, resulting in several limitations. For instance, the assessment criteria may be inconsistent, making it difficult to compare results across different vehicle models. Evaluation outcomes can also be influenced by personal preferences, focus areas, and the expertise of individual experts. Furthermore, expert opinions may be inconsistent or contradictory, and the evaluation process is often cumbersome, requiring the coordination of many experts, which increases costs. Therefore, it is critical to establish an objective, comprehensive, and effective evaluation system for ICLM. However, due to the complexity and diversity of intelligent cockpit applications, formulating such an evaluation system poses significant challenges.

**Challenges in Data Collection and Processing:** Intelligent cockpits involve various sensors and user interaction data, and the diversity and quality of this data directly impact the accuracy and reliability of the evaluation system. Collecting high-quality and representative training and testing datasets is highly complex. In addition, privacy and data security concerns cannot be ignored, especially when handling personal and biometric data. Strict adherence to data protection laws further complicates data collection and processing. Moreover, ICLM must cover a wide range of driving scenarios and user groups to ensure that the evaluation system is comprehensive and applicable. However, actual data collection often faces limitations.

**Model Complexity and Transparency:** ICLM typically involves multi-modal data processing and complex algorithm structures. This complexity makes it difficult to establish standardized and unified evaluation criteria. Since many models are considered "black boxes," understanding and explaining their decision-making processes is challenging, creating higher demands for transparency and interpretability in evaluation standards. In particular, the performance of mainstream models in vehicle scenarios can vary, as some models are not specifically designed for in-car use. Thus, the model's applicability and optimization should be carefully considered during evaluation.

**Diversity in User Experience:** The diverse expectations and needs of different users pose another challenge to the evaluation system. The subjective nature of user experience and individual differences make it difficult to establish a uniform set of evaluation criteria. The system must balance subjective evaluations with objective measurements, which typically require long periods of observation. This can make it difficult to engage actual users in the measurement process.

**Dynamic Changes in Technology and Standards:** The rapid development of intelligent cockpit technology, with new features and innovations emerging constantly, requires that the evaluation system be adaptable and able to keep pace with changes. Currently, the lack of unified industry standards and guidelines results in varying evaluation methods and criteria across different manufacturers and research institutions, which compromises the comparability and consistency of evaluation results.

**Complexities of Real-World Application Environments:** The performance of intelligent cockpits may differ across various driving conditions and usage scenarios, making it essential to account for these complexities in the evaluation system. Cockpits must make real-time decisions and respond to dynamically changing environments, which places high demands on the real-time and adaptive capabilities of the evaluation framework. Some cockpit functions can only be fully realized in specific complex scenarios, adding difficulty to the comprehensive testing and assessment across all possible use cases.

**Selection of Evaluation Metrics:** It is necessary to balance breadth and depth when selecting evaluation metrics. Since ICLM spans multiple domains and dimensions, the evaluation criteria must cover all aspects comprehensively and provide a detailed assessment of each function and feature. However, the diversity of environments, time, and test subjects complicates the measurement process. Different environmental conditions, extended measurement periods, and individual differences among test participants all propose higher requirements for scientific rigor and objectivity in evaluation.

**Challenges in Measuring Long-Term Memory:** Measuring the long-term memory function of intelligent cockpit systems presents its own challenges. Long-term memory is a key factor in evaluating user preferences and behavioural patterns, yet conducting long-term monitoring and evaluation during real-world driving is difficult. This makes it challenging to provide a comprehensive and precise assessment of the cockpit's long-term memory capabilities.

In summary, establishing an objective, comprehensive, and effective evaluation system for ICLM faces limitations in various aspects, including data collection and processing, model complexity and transparency, diversity in user experience, dynamic changes in technology and standards, complexities in real-world applications, the selection of evaluation metrics, and measuring long-term memory.

## 3. Evaluation System for ICLM：P-CAFE

In the context of intelligent cockpits, the ability to understand both the internal and external environments of the vehicle, as well as user intentions, is crucial for optimal decision-making and response execution. This has become a focal point in the development of automotive technology. With the rapid advancement of LLM like ChatGPT and ERNIE Bot, real-time, seamless communication and feedback between users and vehicles through natural language is now a key development focus for enhancing the user experience in automotive design. It means by positioning the LLM as the "brain" of the vehicle's intelligent cockpit, automakers are shifting toward LLM integration.

However, the pace of LLM development has far outstripped academic expectations, leaving gaps in research related to LLM integration, evaluation, risk identification, and human-machine interaction within intelligent cockpits. Stakeholders are still gaining a full understanding of the potential of LLMs in automotive cockpits. Moreover, ICLM are greatly expanding the range of in-car applications. In these complex scenarios, evaluating the performance and user experience of the next generation of ICLM across different contexts has become an urgent issue. Therefore, the aim of this paper is to summarise the relevant research issues and propose a preliminary evaluation system for the assessment of the application of ICLM based on the proposed difficulties and risks for the reference of academia and relevant stakeholders in the industry.

### 3.1 Principles for Selecting Evaluation Metrics

When transitioning from evaluating human-to-agent interaction to evaluating ICLM, the evaluation of in-car LLM-Agent must account for specific constraints related to the human-vehicle-environment interaction. Considering the multi-modal data processing capabilities of LLM-Agents, as well as the close cooperation between users and LLM-Agents (as shown in Figure 2), except for evaluating the ability of LLM-Agent in perception, cognition and decision-making based on previous research on intelligent agents, it is also an indispensable part of the evaluation of the intelligent cockpit's big model that the development evaluation based on the big model and the evaluation based on the user's feedback will be carried out. To this end, the following principles guide the selection of evaluation metrics.

**Purpose-Driven Principle:** The evaluation system should be designed to align with the goal of assessing the capabilities of ICLM. It should consist of typical indicators representing each component of the human-vehicle-agent system, reflecting the overall capability of ICLM from multiple angles.

**Scientific Principle:** The structure of the evaluation system, the selection of indicators, and the derivation of formulas should all be scientifically based. Adhering to this principle ensures the reliability and objectivity of the information obtained, making the evaluation results credible.

**Systematic Principle:** The evaluation system should encompass all relevant aspects of ICLM, forming a comprehensive evaluation framework.

**Independence Principle:** Some factors may be highly correlated, and including all of them could result in redundant indicators, potentially skewing the evaluation results. In such cases, selecting the most representative indicator is preferred, and correlation analysis tools can be used to assess relationships between factors.

**Operability Principle:** The evaluation system should have a clear and logical structure. Indicators should be well-defined, straightforward, and easily quantifiable, ensuring that accurate and quick data can be obtained to facilitate subsequent evaluations.

**Effectiveness Principle:** The system should not only reflect current conditions but also track changes over time to identify problems and take preventive measures. Additionally, it should evolve in response to changing societal values to remain relevant.

**Representativeness Principle:** While the system should cover a broad range of factors, it should prioritize the most significant and critical safety issues. Focusing on too many variables may dilute attention to the key factors, so representative indicators that reflect the overall safety status should be selected.

**Combination of Qualitative and Quantitative Analysis**: The evaluation system should integrate both qualitative and quantitative analyses. Quantitative data allows for more accurate insights into the system, while qualitative factors can be scored and quantified through expert opinions when statistical data is lacking.

### 3.2 Construction of P-CAFE

To evaluate the capabilities and user experience of ICLM, five primary dimensions: Perception, Cognition, Action, Feedback, and Evolution were selected as an Evaluation Indicator System for ICM, called ***P-CAFE***. These were established as the first-level indicators of the evaluation system. In collaboration with users, automotive engineers, and experts from the automotive and large model fields, second- and third-level indicators were selected and defined according to the principles of indicator selection.

A human driver's successful task execution involves the coordinated operation of many basic psychological processes, such as perception, attention, memory, decision-making, and motor actions. Understanding human interaction with (partially) automated vehicles must be placed in the context of human driving behaviour, and must also include processes added due to the introduction of automation, such as monitoring the automated system's behaviour and status, predicting its actions, and building trust. While automated driving technology has advanced significantly, intelligent cockpits equipped with large models still follow this framework.

ACT-R is a cognitive architecture that provides a unified theory of human cognition [12]. One notable feature of this architecture is that it explains how memory, perception, attention, and motor skills interact as a coherent system to produce cognition and behavior. Cognitive architectures can be extended to complex tasks related to human-computer interaction, such as driving, managing autonomous vehicles, and aviation [13]. Since there already exists a form of intelligent driving—human driving—it is reasonable to draw inspiration from human cognition and intelligence when designing other forms of "vehicle intelligence." Cognitive computing is an AI-based system that interacts with humans like a human would, interprets context, analyses past user behaviour, and infers based on the interaction. Cognitive computing assists humans in decision-making, while AI-based systems aim to make better decisions on behalf of humans.

Human cognition has long been a central focus of a broad field of research Based on the two-dimensional conceptual space proposed by Poirier and Chicoisne for categorising cooperation between two entities in a cognitive system (shown in Figure 2.) [14], Plebe et al. describe a collaborative interaction paradigm between a human and an AI agent by invoking the rider-horse metaphor, also called H-Metaphor [15]. In this framework, the roles of the agent and user vary in terms of perception, decision-making, and execution, similar to the relationship represented in quadrant 2 of Figure 2.

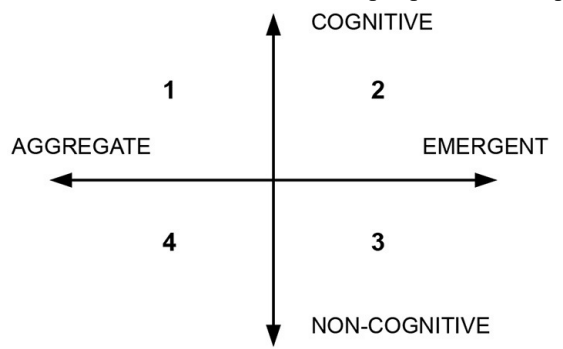

Figure 2 Conceptual Space Classifying Relationships Between Two Entities in a Cognitive System [14]

In the context of intelligent cockpits, perception involves inputting multiple data types, including data received from general visual models, voice models, and multimodal models. From a user-vehicle (LLM-Agent)-environment perspective, the system's perception of sound inputs, such as wake-up calls and command recognition, is critical and central to the user experience [16-17]. Another important input source is the recognition of images inside and outside the vehicle [18]. Apart from the vehicle's hardware-based data perception, large models have the ability to access multimodal information from the internet in real-time. This capability, as described in section 1.1, is a key characteristic distinguishing third-generation intelligent cockpits from second-generation ones, enabling various application ecosystems such as travel and smart home ecosystems [19]. Therefore, auditory perception, visual perception, and ecological connectivity are selected as second-level indicators for evaluating the perception capabilities of P-CAFE.

The advantages of Intelligent Cockpit powered by a big model over a general Intelligent Cockpit are the ability to process multimodal information input and the ability of the big language model to process natural language as described in Section 1.1, which significantly enhance the naturalness and intelligence of human-vehicle interaction. In the car, the Agent generated based on the big model can not only realise simple voice interaction based on natural language, but also, from the perspective of computer cognition, make the Agent "think" and "reason" like a human being through the input of natural language information, and have advanced cognitive functions, so that the big model can carry out limited reasoning according to the acquired information and the knowledge of the Internet [20]. As described in sections 1.2 and 1.3 more advanced cognitive capabilities enable the agent to provide personalized and human-like emotional and intent recognition through monitoring and data collection of user-vehicle-environment states [21-22]. Thus, natural language processing, reasoning based on large knowledge bases or datasets, and real-time intent recognition are considered second-level indicators for evaluating cognitive capabilities in P-CAFE.

In cognitive theory, besides perception and cognitive reasoning, decision-making is also an important and fundamental issue, as well as a starting point for planning the next behavioural steps regarding the user and ICLM, and guiding the execution of subsequent actions [23]. For ICLM, real-time decision-making based on changing scenarios and vehicle status, as well as the allocation of tasks and decision authority in human-machine collaboration, are key second-level indicators under the "action" dimension [24]. In the automotive field, quick and accurate decision planning, command response and execution are essential to making vehicle interaction more natural and intelligent, which is why they are considered other critical second-level action indicators for P-CAFE [25].

The feedback dimension of user experience in intelligent cockpits has always been a focus of human-computer interaction and human factors engineering throughout the three generations of cockpit development. Human-Machine Interface (HMI) design has long been a core design element in automotive design, often evaluated by interface usability [26]. After the integration of large models into the cockpit, the user's trust in the technology has become a key research area in AI. It is also a characterisation of the user's trust in the large model of the Intelligent Cockpit during actual use, and research in recent decades has resulted in more mature evaluation methods being used [26]. The ability of large models to process multimodal interactions, such as visual, auditory, touchscreen, and voice signals, brings convenience but also affects user load and experience [27-28]. Monitoring user load is an integral part of driver state monitoring systems, developed by OEMs as key projects [29]. In addition to monitoring driver load, detecting changes in emotional states is also a core function. Emotional assessment has been a frequent topic in evaluating large models. Therefore, usability, trust, user load, and emotional feedback are selected as second-level indicators in the feedback dimension for P-CAFE [30].

The ability to self-evolve is a key advantage of large models compared to earlier AI models [27]. As mentioned in section 1.1, they can perform unsupervised learning using environmental information and contextual cues [31]. For the LLM-Agent in the Intelligent Cockpit, scholars have proposed a new paradigm for multimodal general agents in intelligent cockpits, as shown in Figure 3[9]. This framework comprises five modules:

1) environment and perception (task planning and skill observation),

2) agent learning,

3) memory,

4) agent action,

5) cognition.

In this framework, LLM-based agents can continuously evolve through self-learning and self-assessment. Hence, memory and learning ability are considered second-level indicators under the evolution dimension for P-CAFE.

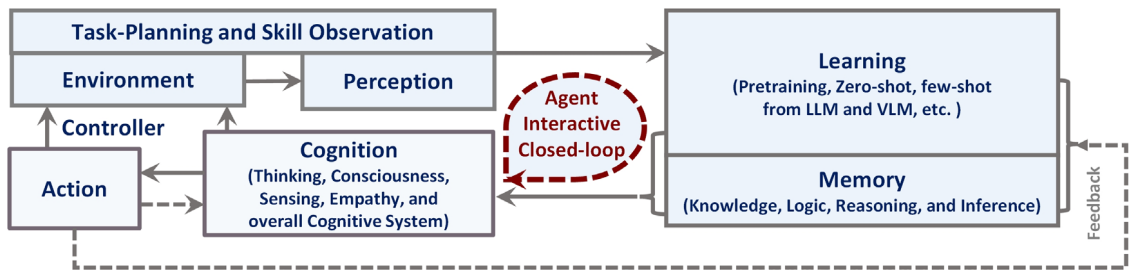

Figure 3 New Paradigm of Multimodal General Agents

### 3.3 Measurement of Evaluation Metrics

The evaluation of P-CAFE is conducted based on user interaction tasks, primarily through voice dialogue. The feedback from individual interaction tasks serves as the smallest unit of analysis. A comprehensive evaluation is obtained through user experience assessment methods [32], including rating scales, combined with eye-tracking experiments and EEG (electroencephalogram) testing [33]. The evaluation metrics can be categorized based on the type of intelligent cockpit tasks, which include measurable metrics such as accuracy and efficiency, as well as subjective metrics that are difficult to quantify, which pertain to user experience.

**Experimental Environment.** The evaluation of the automotive large model should be carried out in a quiet testing environment, with continuous monitoring of environmental noise. The sound pressure level of ambient noise near the microphone should be between 45 and 65 dB(A), with a signal-to-noise ratio greater than 15 dB. It is recommended to use at least two video recording positions and at least one audio collection point. The distance between the recording device and the sound source should be 35–55 cm (overhead) and 65–75 cm (dashboard). The camera setup should include one to record the interface of the vehicle screen and at least one other to capture the interior environment of the vehicle.

**Measurement Methods.** The evaluation of P-CAFE involves not only the analysis of multimodal data but also the process of human-computer interaction. The factors required for this analysis generally reflect five types of indicators:

(i) Quantity, (ii) Diversity, (iii) Speed, (iv) Accuracy, and (v) Value [34]. The measurement of the model's indicators can be categorized accordingly.

**Accuracy Indicators**: These are assessed by measuring the correctness of the large model's feedback to a given number of questions or its alignment with human-set standards [35].

**Efficiency Indicators**: Metrics such as task execution efficiency and response speed are evaluated by recording the experiment process at 60 frames per second (fps) or higher. The model's interaction efficiency is calculated based on frame count.

**Subjective Indicators**: Various ergonomic and psychological scales are used to obtain effective feedback from users regarding the large model in the intelligent cockpit. These include the NASA-TLX cognitive load scale [36] and the PANAS-SF emotional analysis scale [37], which measure usability, trust, workload, and emotional dimensions of user perception.

**Objective Indicators**: Physiological and biological data can be collected using non-invasive wearable devices. Devices such as wristbands, head-mounted eye trackers, EEG machines, and electromyography (EMG) equipment can measure users' postures, eye movements, brain activity, and muscle signals to assess fatigue, cognitive state, and emotional changes. Additionally, more precise biological information can be obtained by analyzing the user's sweat or saliva samples.

### 3.4 Evaluation Methods for P-CAFE

To better study vague phenomena, fuzzy mathematics was introduced. Both the Analytic Hierarchy Process (AHP) and Fuzzy Comprehensive Evaluation (FCE) are comprehensive evaluation methods based on fuzzy mathematics. The Fuzzy Analytic Hierarchy Process (FAHP) emerged to address the shortcomings of both methods.

**Analytic Hierarchy Process (AHP)**

Proposed by American operations researcher SAATY, AHP breaks down a complex problem into several ordered levels and factors, creating a hierarchical structure model, as shown in Figure 4. This method ultimately forms a multi-objective decision-making framework for simpler problems. In practice, experts are consulted to compare the importance of various factors within each level. These comparisons form a matrix, from which the maximum eigenvalue and eigenvector are calculated, determining the weights of the factors.

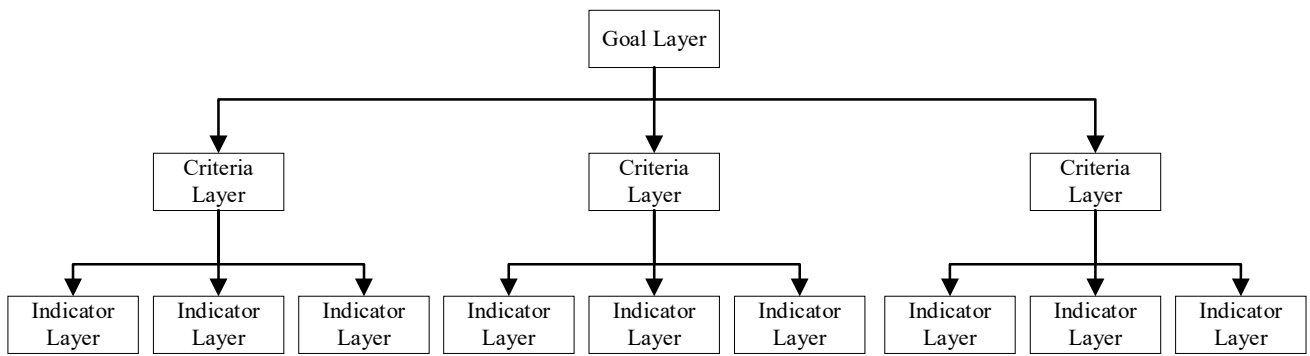

Figure 4 Hierarchical structure model of AHP

To determine the weight coefficients of each indicator, a judgment matrix for the lower level relative to the upper level must be established. This matrix is reciprocal, meaning that if factor $A_i$ is compared with $A_j$, the matrix $A = (a_{ij})_{n*n}$ which element satisfies:

$$a_{ij} = \frac{1}{a_{ji}},$$

$$a_{ij} > 0,$$
$$a_{ij} = 1,$$

The judgment matrix can be represented as:

$$\begin{bmatrix} a_{11} & a_{12} & \cdots & a_{1j} \\ \vdots & & \ddots & \vdots \\ a_{i1} & a_{i2} & \cdots & a_{ij} \end{bmatrix}$$

The judgement matrix is based on the judgement criteria of an element at the upper hierarchy level, and the invited experts use quantitative assessment scales based on their daily work experience to compare one indicator factor with another at the same hierarchy level and assign values to the elements of the judgement matrix. The assessment criteria can be classified into nine levels according to their relative importance, with five basic criteria and four intermediate criteria, and the significance of each criterion is shown in Table 2.

Table 2  1-9 Scale significance statements

| Scale | Statements |
| --- | --- |
| 1 | Two factors contribute equally to the objective. |
| 3 | One factor is moderately important than the other. |
| 5 | One factor is strongly important than the other. |
| 7 | One factor is very strongly important than the other. |
| 9 | One factor is extremely important than the other. |
| 2, 4, 6, 8 | Used to represent intermediate importance between the above values. |

The determination of indicator weights is based on using the judgment matrix to measure the importance of factors related to a certain element in the higher level from the lower level. For the constructed judgment matrix $A = (a_{ij})_{n*n}$, its largest eigenvalue $\lambda_{max}$ and eigenvector $W = (W_1, W_2, \ldots, W_n)^{\mathrm{T}}$ are calculated to determine the evaluation indicator weights.

$$\overline{W_i} = \sqrt[n]{\prod_{j=1}^{n} a_{ij}} \quad i,j = 1,2,\ldots,n$$

$$W_i = \frac{\overline{W_i}}{\sum_{i=1}^{n} \overline{W_i}}$$

$$\lambda_{max} = \frac{1}{n} \sum_{i=1}^{n} \frac{(AW)_i}{W_i}$$

Finally, to ensure the reliability of the results, a consistency ratio (CR) is used to test the consistency of the judgment matrix. The consistency is measured by the value of CR, with a general threshold of 0.1. When CR is greater than 0.1, it indicates that the matrix is inconsistently constructed and needs to be adjusted; when CR is less than 0.1, it indicates strong consistency and that the matrix is reasonably constructed. The formula for calculating the CR value is as follows:

$$CR = \frac{CI}{RI}$$

$$CI = \frac{\lambda_{max} - n}{n - 1}$$

Where the RI value can be retrieved from the Random Consistency Index (RI), as following Table 3.

Table 3 The Random Consistency Index (RI) table

| order n  | 1 | 2 | 3    | 4   | 5    | 6    | 7    | 8    | 9    | 10   | 11   |
|----------|---|---|------|-----|------|------|------|------|------|------|------|
| RI value | 0 | 0 | 0.58 | 0.9 | 1.12 | 1.24 | 1.32 | 1.41 | 1.45 | 1.49 | 1.51 |

**Fuzzy Comprehensive Evaluation (FCE)**

Based on fuzzy mathematics as the theoretical foundation, the unquantifiable and ambiguous influencing factors within the research subject are treated as a set. Using the theory of membership degree, corresponding membership functions are constructed, and the evaluation results are presented in the form of fuzzy sets. The specific steps are detailed below in Figure 5. The fuzzy comprehensive evaluation method is advantageous in its ease of understanding, ease of application, and rational evaluation criteria. When addressing complex, multi-factor, and multi-level issues, a fuzzy comprehensive evaluation is more objective compared to other mathematical methods.

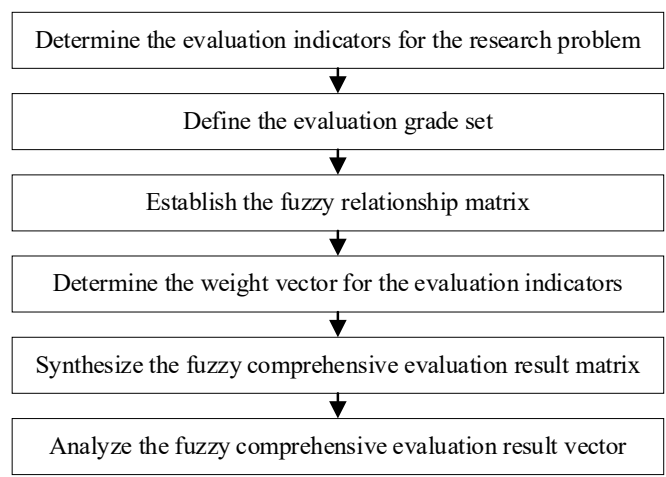

Figure 5 Specific steps of the fuzzy comprehensive evaluation method

**Fuzzy Analytic Hierarchy Process (FAHP)**

The Fuzzy Analytic Hierarchy Process (FAHP) is an evaluation method that combines the characteristics of both the Analytic Hierarchy Process (AHP) and Fuzzy Comprehensive Evaluation (FCE). Its operational steps are essentially the same as AHP, but with a key difference: while AHP typically constructs a judgment matrix by comparing factors in pairs, FAHP uses pairwise comparisons to construct a fuzzy consistency judgment matrix that better reflects human thought processes. This approach better captures the ambiguity of real-world problems and helps reduce the influence of subjective factors during analysis.

Compared to AHP and FCE, FAHP builds on AHP's approach to solving complex qualitative problems layer by layer and integrates fuzzy mathematical principles to determine the weights of fuzzy criteria. By combining these two methods, FAHP mitigates the negative effects of individual expert opinions on system scoring, making the data collection and analysis process more flexible and the evaluation results more accurate and realistic. The general steps of FAHP are as follows:

1.  **Determine the Factor Set**

The collection of influencing factors is referred to as the "factor set" and is generally represented by U: $U = (U_1, U_2, U_3, ..., U_n)$

2.  **Build the Hierarchical Structure**

The influencing factors are broken down layer by layer to form a hierarchical evaluation structure model from bottom to top.

### 3. Construct the Fuzzy Judgment Matrix for Each Layer

Expert surveys are conducted, and the results are compared pairwise to determine the relative importance of influencing factors at the same level. This helps build the fuzzy judgment matrix, represented as $A = (a_{ij})_{n*n}$. The 0.1-0.9 scale method is commonly used for comparison, as shown in Table 4.

A fuzzy reciprocal matrix is achieved when the elements of matrix $A = (a_{ij})_{n*n}$ meet the following conditions:

$$a_{ii} = 0.5, \quad i = 1, 2, \ldots, n$$
$$0 \leq a_{ij} \leq 1, \quad a_{ij} + a_{ji} = 1, \quad i, j = 1, 2, \ldots, n$$

Table 4 Explanation of the 0.1-0.9 Scale for Importance

| Judgment Scale | Explanation |
|---|---|
| 0.5 | Equally important |
| 0.6 | Slightly more important |
| 0.7 | More important |
| 0.8 | Significantly more important |
| 0.9 | Extremely important |
| | Opposite comparisons |
| 0.1, 0.2, 0.3, 0.4 | When element $a_i$ is compared to the element $a_j$, yielding a judgment $r_{ij}$, then when the element $a_j$ is compared to the element $a_i$, the resulting judgment is $r_{ji} = 1 - r_{ij}$. |

### 4. Consistency Check of the Fuzzy Judgment Matrix and Conversion to Fuzzy Consistency Matrix

When comparing two factors, subjective judgment and problem complexity can affect the results. Therefore, it is necessary to check for consistency in the judgment matrix. If the elements $a_{ij}$ of matrix $A$ satisfy the equation below, the matrix is termed the fuzzy consistency matrix $R$:

$$a_{ij} = 0.5 + a_{ik} - a_{jk}, \quad i, j, k = 1, 2, \ldots, n$$

The fuzzy consistency matrix $R = (r_{ij})_{n*n}$ is obtained by summing each row of the fuzzy reciprocal matrix $A = (a_{ij})_{n*n}$ and transforming it mathematically:

$$r_i = \sum_{k=1}^{n} a_{ik}, \quad i = 1, 2, \cdots, n$$

$$r_{ij} = \frac{r_i - r_j}{2(n-1)} + 0.5$$

### 5. Weight Calculation and Ranking

Using the weight formula, calculate the weights $W_i = (\omega_1, \omega_2, \ldots, \omega_n)^T$ and the largest eigenvalue

$$\overline{W_i} = \sqrt[n]{\prod_{j=1}^{n} r_{ij}} \quad i, j = 1, 2, \ldots, n$$

$$\omega_i = \frac{\overline{W_i}}{\sum_{i=1}^{n} \overline{W_i}}$$

$$\lambda_{max} = \frac{1}{n} \sum_{i=1}^{n} \frac{(AW)_i}{W_i}$$

Additionally, weights can be calculated using a linear function based on transformation coefficients. The specific formula is:

$$\omega_i = \frac{1}{n} - \frac{1}{2\theta} + \frac{1}{n*\theta} \sum_{k=1}^{n} r_{ik}, \quad i = 1, 2, \cdots, n, \ 0 \leq \theta \leq 0.5$$

Consistency should be checked using the consistency ratio (CR) to ensure the matrix is reasonable. If *CR > 0.1*, the

matrix needs adjustment. If *CR < 0.1*, the matrix is considered consistent. The CR formula is:

$$CR = \frac{CI}{RI}$$

$$CI = \frac{\lambda_{max} - n}{n - 1}$$

The RI values are obtained from the Random Consistency Index (RI) table (Table 3 ).

6. **Determine Membership Degrees and Evaluation Matrix**

If the membership degree of the *i*-th element in the factor set *U* to the first element in the evaluation set *V* is denoted as $z_{i1}$, then the single-factor evaluation result for the *i* -th element can be expressed as a fuzzy set:

$$Z_i = (z_{i1}, z_{i2}, \dots, z_{in})$$

The *n* dimension single-factor evaluation sets $Z_1, Z_2, \dots, Z_n$ are arranged as rows to form the matrix $Z_{n \times m}$, which is referred to as the fuzzy comprehensive evaluation matrix.

$$z_{ij} = \frac{\text{Number of evaluations at level}}{\text{Total number of experts}}, \quad i = 1,2,\cdots,n, \quad j = 1,2,\cdots,m$$

7. **Obtain Single Layer Factor Evaluation Results**

Calculate the single-factor evaluation vector $B_i$ and the evaluation score $S_i$. The maximum membership method is used to determine the results:

$$B_i = W_i * Z_i$$
$$S_i = B_i * V$$

8. **Overall Evaluation and Conclusion**

The evaluation vectors $B_i$ for each single factor are arranged as rows to form the overall evaluation vector $B_{total}$. The weights of the primary indicators are combined to form the weight vector *W*. The overall evaluation vector BBB for the overall objective is then calculated using the following formula, and the result is determined using the maximum membership degree method:

$$B = W * B_{total}$$

The comprehensive score for the overall objective is obtained using the following formula:

$$S = B * V$$

Where *V* represents the evaluation set, and *S* is the final overall score.

4. **Conclusion**

This study aims to evaluate the capabilities of ICLM. Based on the current development and integration of intelligent cockpits and large language models, and considering the influencing factors and limitations from the fields of computer cognition and automotive human-machine interaction, an evaluation indicator system, P-CAFE, and method for assessing the capabilities of large models in intelligent cockpits was developed. With the rapid advancement of large model technology, the capabilities of models integrated into intelligent cockpits are iterating quickly, and the definition and composition of these models will continue to evolve. Ongoing practice and research will require dynamic updates and further in-depth exploration. Table 5 presents the five primary indicators, along with their corresponding secondary indicators and explanations of P-CAFE. The weight of each indicator was determined according to the method outlined in Section 3.3, based on the evaluation and review by 10 experts.

Table 5  Overall Indicator System of P-CAFE

| Primary Indicator | Secondary Indicator | Explanation of Secondary Indicator |
|---|---|---|
| **Perception** | Auditory Perception | Evaluation of the model's ability to perceive sound |
| | Visual Perception | Evaluation of the model's ability to perceive visual stimuli |
| | Ecological Connectivity | Evaluation of the model's ability to connect with information from transportation, daily life, mobile devices, smart home systems, and real-time internet searches |
| | Multimodal Input | Evaluation of the model's capability to handle multiple modes of input |
| **Cognition** | Natural Language Processing | Evaluation of the model's accuracy in recognizing natural language, the maximum length of input it can process, and its ability to perform logical reasoning and semantic inference in everyday conversations, English, minor languages, and informal grammar |
| | Knowledge Reasoning | Evaluation of the model's ability to respond to queries about in-car manuals, driving knowledge, and cultural and ethical matters |
| | Intent Recognition | Evaluation of the model's accuracy in recognizing user commands from different seat areas, including those requiring contextual understanding, personal preferences, and situational needs. Also evaluates its ability to detect and filter non-command or unreasonable input |
| **Action** | Decision-Making | Evaluation of the model's ability to dynamically adjust decisions based on real-time changes in scenarios and vehicle status, and its capacity to distribute tasks and decision-making authority during human-machine collaboration |
| | Planning | Evaluation of the model's comprehensive planning ability for decision-making, including its ability to judge the match between planning schemes and decision goals, and metrics such as the number of plans and average planning time |
| | Execution | Evaluation of the model's ability to execute planning tasks, including its execution success rate across different seating areas, execution speed, and the number of modes used during execution |
| **Feedback** | Usability | Evaluation of the model's ability to effectively, safely, and efficiently support users in completing specific tasks, while ensuring user satisfaction, ease of learning, memorability, and efficiency. The International Organization for Standardization (ISO) defines usability as the effectiveness, efficiency, and satisfaction with which a product can be used by specific users for specific purposes in a given context[29] |
| | Trust | Evaluation of users' confidence and expectations regarding the model's reliability, safety, predictability, and transparency in various driving scenarios |
| | Load | The total cognitive and physical demands and stress users experience when interacting with the model |
| | Emotion | Evaluation of the model's ability to convey and manage emotional information through various means, such as human-like expressions, role-playing, emotional responses, and emotional regulation during user interactions |
| **Evolution** | Memory | Evaluation of the model's ability to retain and utilize contextual information over one or multiple conversation rounds |
| | Learning | Evaluation of the model's ability to dynamically adjust its functions and services by continuously monitoring and analyzing user behavior and preferences, thereby improving its own characteristics over time |
| | Personality | Evaluation of the model's display of consistent behavior patterns, attitudes, preferences, and emotional characteristics during user interactions |

**Funding**

This work was supported by the research project on Modelling and Evaluation of User Behaviour Analysis Based on Smart Cockpit Interaction Multimodal Data (L247008).

**Declaration of competing interest**

The authors declare that they have no known competing financial interests or personal relationships that could have appeared to influence the work reported in this paper.

**Data availability**

No data was used for the research described in the article.

**Acknowledgements**

On behalf of all the authors, we would like to thank the ten experts who participated in the evaluation.